\documentclass{achemso}
\usepackage[T1]{fontenc}       
\usepackage[section]{placeins}
  
\title{Thermodynamics of the Oxygen Evolution Electrocatalysis in Metal-Organic Frameworks}

\author{Terence Musho} 
\affiliation{Department of Mechanical and Aerospace Engineering, West Virginia University, Morgantown, WV 26506-6106, USA.} 
\email{tdmusho@mail.wvu.edu}
\author{Jiangtan Li}
\affiliation{Department of Mechanical and Aerospace Engineering, West Virginia University, Morgantown, WV 26506-6106, USA.} 
\author{Nianqiang Wu}
\affiliation{Department of Mechanical and Aerospace Engineering, West Virginia University, Morgantown, WV 26506-6106, USA.} 

\begin{document}

\begin{abstract}
Metal-organic frameworks (MOFs) provide a versatile and tailorable material platform that embody many desirable attributes for photocatalytic water-splitting. The approach taken in this study was to use Density Functional Theory (DFT) to predict the thermodynamic energy barriers of the oxygen evolution reaction (OER) for three MOF functionalizations. A Zr-MIL-125 MOF design was selected for this study that incorporates three linker designs, a 1,4-benzenedicarboxylate (BDC), BDC functionalized with an amino group (BDC+NH$_2$), and BDC functionalized with nitro group (BDC+NO$_2$). The study found several key differences between homogeneous planar catalyst thermodynamics and MOF based thermodynamics, the most significant being the non-unique or heterogeneity of reaction sites. Additionally, the funcationalization of the MOF was found to significantly influence the hydroperoxyl binding energy, which proves to be the largest hurdle for both oxide and MOF based catalyst. Both of these findings provide evidance that many of the limitations precluding planar homogeneous catalysts can be surpassed with a MOF based catalyst. While none of the MOF designs selected for this study out-performed state-of-the-art oxide based catalysts, the BDC+NH$_2$ proved to be the best with a predicted over-potential for spontaneous OER evolution to be 3.03eV.
\end{abstract}

\section{Introduction}
The viable production of hydrogen from renewable sources has proven to be a difficult feat but the success of such technology boasts a tremendous effect to alleviate dependence on fossil fuels~\cite{lewis06,hurst10}. One of the more direct and promising routes for the sustainable generation of hydrogen through water splitting is the utility of solar energy~\cite{turner08,turner04,bolton96,lewis07}. Granted the production of hydrogen from water can be catalyzed through external electrical potential~\cite{stojic03} and thermal reformers~\cite{nakamura77}, however, the energy required by many of these processes is often much greater than the resulting energy of the hydrogen products. Thus, in evaluating the overall energy balance there is an negative sum of energy production when using these conventional hydrogen production techniques and materials. One of the more viable and sustainable approaches to hydrogen production is to harness radiation from the Sun to drive the low temperature electrolysis of water splitting. Unfortunately, off-the-shelf materials are not readily available that can be directly applied to the low-temperature photocatalytic conversion of water to hydrogen. While many approaches have focused on the improvement of planar cathode (limiting electrode for water splitting) there has not been much deviation beyond these materials. Recent theoretical studies~\cite{rossmeisl11} of the oxygen evolution reaction (OER) on homogeneous materials has revealed fundamental limits governed by the physics of the hydroxyl and hydroperoxyl binding. One method to circumvent these physical limitation is to select a material with heterogeneous reaction sites that are situated near each other. In satisfying this requirement, the material selected for this study is a metal-organic framework (MOF) material. Metal-organic frameworks are in the most general description a ordered arrangement of metal clusters or metalloids connected by an organic ligand or linker. The tailorability of these materials come in the selection of not only the inorganic metalloid but also the functionalization of the organic ligands. One of the major obstacle that has limited a MOF based photocatalyst solution for water splitting lies in designing a optimal material in a extremely large design space. This study provides a computational method for exploring this designing space with a specific concentration on predicting the thermodynamics of the oxygen evolution as a function of organic ligand functionalization.
\begin{figure}[!ht]
\includegraphics[width=1.0\columnwidth]{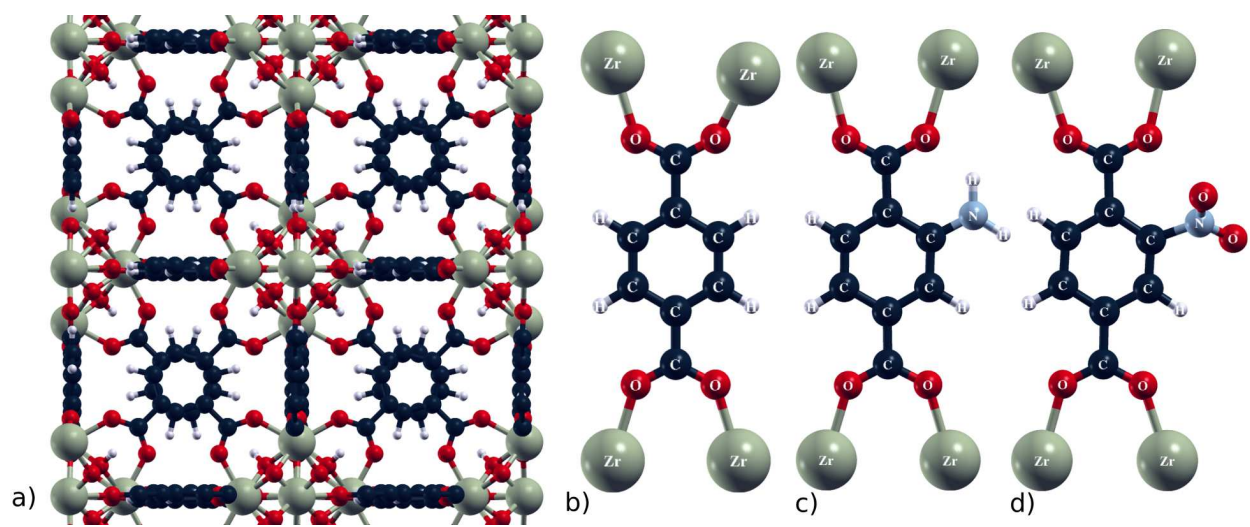} 
\caption{Illustration of a) the unit cell and the three primitive unit cells of the studied metal organic framework material. Each MOF framework is based on Zr-UiO-66 design~\cite{valenzano11} and the linker design is b) benzenedicarboxylate (BDC), c) BDC functionalized with NH2 (BDC+NH2), and d) BDC functionalized with NO2 (BDC+NO2).}
\label{f:uc}
\end{figure}
In developing a material for photocatalytic water oxidation the material must exemplify two important attributes, 1) the material must readily absorb solar radiation while minimizing phonon production, and 2) the material must readily catalyze water through the oxygen evolution reaction. In this study, it is hypothesized that a MOF could embody both of these attributes~\cite{lee09,peng13} while controlling their attributes through careful selection of an appropriate functional group attached to the ligands. Previous studies~\cite{musho14,lin12} have demonstrated that a ligand of the MOF material is the photoactive portion of the material. The ligand can be functionalized for optical absorption and near full utility of solar radiation. With these previous studies in mind, this study will focus on the latter attribute by investigating the thermodynamics of the OER in response to the ligand functionalization. 

The photocatalytic water-splitting conversion for hydrogen production is a demanding reaction. In a photocatalytic cell, the anode, which involve the hydrogen evolution reaction proves to be less demanding than the cathode electrode, which is responsible for the oxygen evolution reaction. The cathode, or more specifically, the oxygen evolution reaction involving two half reactions with each half reaction involves two reaction steps~\cite{hurst10}. The cathode, in this study, the MOF material, needs to favorable selectivity towards a range of radicals in both of these half reactions. Furthermore, there is a range of other attributes that are important for a high performing catalyst. These attributes are in the absences of any kinetic consideration and this study simply emphasizes the thermodynamic attributes. The limiting thermodynamic attributes can be distilled into two attributes, 1) the high activation energy of the oxygen evolution reaction (OER) and 2) a high areal density of reaction sites~\cite{canepa13}. As eluded to in the previous paragraph, MOF materials provide a material platform that can not only be tailored from the perspective of selectivity, but also the porosity, through the selection of ligands~\cite{yang11}. The approach taken in this study is adapted from previous successful predictions of the oxygen evolution reaction thermodynamics of planar cathodes using the computational method of density functional theory (DFT)~\cite{rossmeisl11}. A deviation from these previous studies is the proposed method is the reactions takes place inside of the large pores of the MOF structure opposed to the free surface. Additionally, this study is interested in relative comparison of the functionalization on the thermodynamics of the OER.

\subsection{Material Design}
The specific MOF design was a Zr-MIL-125 MOF~\cite{hui13,long12,hendon13} with an incorporated linker design of 1,4-benzenedicarboxylate (BDC), BDC functionalized with and amino group  (BDC+NH$_2$), and BDC functionalized with nitro group (BDC+NO$_2$). Figure~\plainref{f:uc} provides an illustration of the primitive unit cell of the three MOF designs. The primitive unit cells consist of approximately 114 atoms. A conventional unit cell has approximately 324 atoms. The nominal lattice constant ($a$) for all of the MOF designs is 14.74\AA. The space group is cubic $\bar{4}3m$ and follows the MIL-125 MOF standard. A CIF descriptions can be found in the literature~\cite{banglin09,hendon13}.

\subsection{Computational Details}
A density functional theory (DFT) approach~\cite{qe} was implemented to predict the ground state thermodynamic properties for each of the thermodynamic steps. To reduce the computational expense of the simulations, only a single primitive cell was analyzed for each configuration. In addition, to using the symmetry of the primitive cell, a pseudodized wave function approach was used to reduced the computational expense. The functional form of the pseudowave functions were based on Perdew-Burke-Ernzerhof (PBE) ultrasoft potentials. Several other functionals such as BLYP and their hybrid counter parts were investigated but the PBE was found to be most accurate and stable. The k-point mesh was sampled using a Monkhorst Pack 4x4x4 grid with a offset 1/4,1/4,1/4. To account for the Van der Waals interaction a Van der Waals correction term~\cite{grimme06,barone09} was incorporated, which introduced some empiricism into the calculation. The scaling parameter (S6) were specified to be 0.75 and cut-off radius for the dispersion interaction was 200 angstroms.  Both the ion and unit cell geometries were relaxed to a relative total energy less than 1x10$^{-10}$ and overall cell pressure of less than 0.5kBar.

For this analysis, only the thermodynamics of the reaction steps were of interest and not the kinematics of the reaction. While their are other approach such as nudge elastic band (NEB) theory, which relies on an image potential to determine the reaction pathways, this was reasoned unnecessary for calculation of the thermodynamic energy barriers. Furthermore, to aid in the computational stability, the was reactants were placed near the linker and functional groups and the ions were relaxed. Once these reactants were placed within the unit cell, the whole ensemble was relaxed, including ion position and lattice parameters. Once the structure was relaxed completely, the total energy was recorded. The reader should be made aware that the total energy determined through this method will depend on the pseudo-potential used and caution should taken when reporting these energies. A detailed explanation of the Gibbs calculation can be found in the supplemental material.

The analysis began by determining the total energies of all the constitute atoms within a large box. In addition, the total energy of MOF structure as-is was calculated along with the total energy of the reactants, which include H$_2$O, HO, HO$_2$, H$_2$, and O$_2$, as outlined in Equations~\plainref{eq:1} through Equation~\plainref{eq:4}. The difference between the reactants and the energy of their constituent atoms provides the enthalpies of formation. The reader should note that there is some limitations in the ability of DFT to predict hydrogen bonding due to the over-analytics of the functional. This limits the ability to predict the formation energy of hydrogen (H) from water (H$_2$O) and dihydrogen (H$_2$). In order to account for these inaccuracies, a standard hydrogen electrode potential was specified as 2.46V instead of the calculated DFT determined formation energy of hydrogen. In making this assuption for the hydrogen formation energy the result in turn are referenced to the standard hydrogen electrode.
\begin{figure*}
\centering
\includegraphics[width=1.0\textwidth]{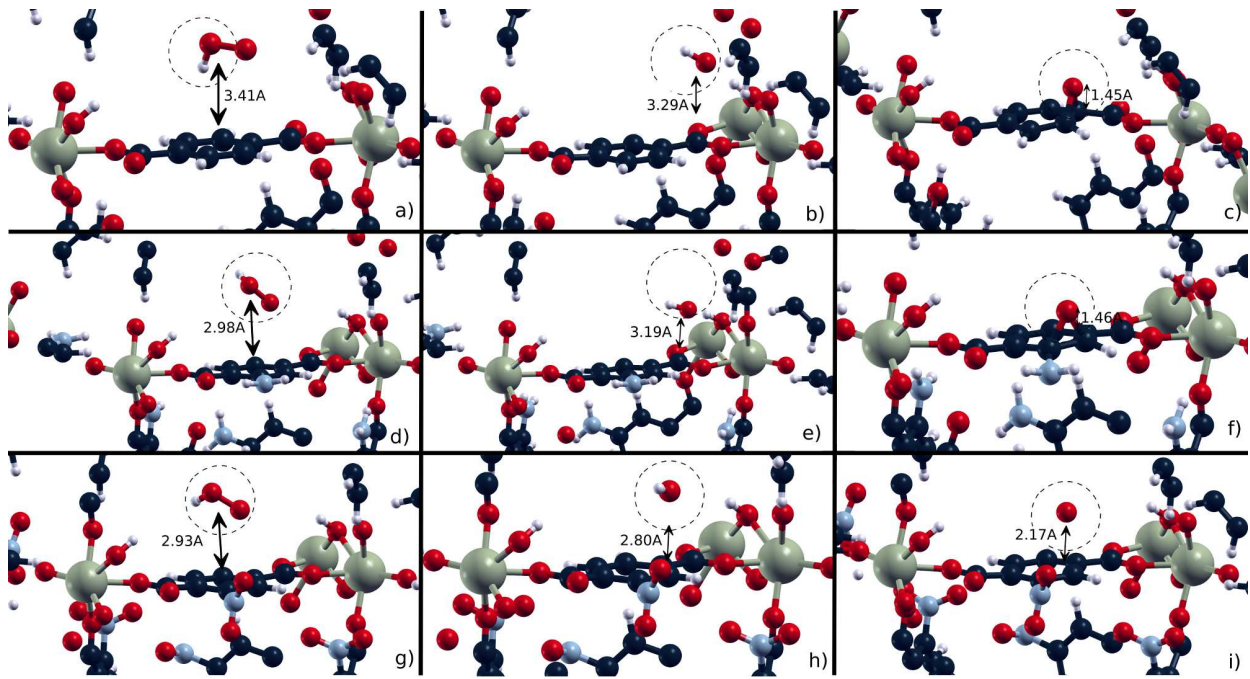} 
\caption{Illustration of the hydrperoxyl (HOO) adsorption (a,d,g), hydroxyl (HO) adsorption (b,e,h), and oxygen (O) adsorption (c,f,i) on the three funtionalized MOF designs. The top figures (a,d,g) is MOF BDC, middle figures (b,e,h) is MOF BDC+NH$_2$, and bottom figure (c,f,i) is MOF BDC+NO$_2$.}
\label{f:atomic_all}
\end{figure*}
\subsection{Thermodynamics of OER}
The oxygen evolution reaction is similar to the hydrogen evolution reaction, which involves a series of single electron charge transfer steps. However, the oxygen evolution reaction involves two half reactions~\cite{hurst10}. These single electron charge transfer steps follow the sequence of Equations~\plainref{eq:1}-\plainref{eq:4}. The asterix ($^*$) denotes the adsorption of the radical to the pore surface of the framework. As will be discussed in the later sections, the absorption site is not the same location for each reaction, as is often the case in homogeneous single crystal planar catalyst. This is an important concern that can be taken advantage of to surpass the physics based limitation imposed by homogeneous planar catalysts.

\begin{equation}
2H_2O \leftrightarrow OH^*+H^++e^-+H_2O
\label{eq:1}
\end{equation}

\begin{equation}
OH^*+H^++e^-+H_2O \leftrightarrow O^*+2H^++2e^-+H_2O
\label{eq:2}
\end{equation} 

\begin{equation}
O^*+2H^++2e^-+H_2O \leftrightarrow OOH^*+3H^++3e^-
\label{eq:3}
\end{equation}

\begin{equation}
OOH^*+3H^++3e^- \leftrightarrow O_2+4H^++4e^-
\label{eq:4}
\end{equation}

The approach taken in this research is to model each of these reaction steps independent of each other. Consideration of the four thermodynamics steps that comprise the complete OER. The term that ultimately governs the spontaneous evolutionary of the OER is the required over-potential to evolve the reaction. The over-potential is governed by the following relation,

\begin{equation}
\eta=\frac{\Delta G}{nF}+dissipation,
\label{eq:diss}
\end{equation}

where $\eta$ is the required over-potential, $F$ is the Faraday constant (94585.Coulomb/mol), and $n$ is the number of moles. The first term is the result of the thermodynamics of the reaction barriers, moreover, the largest reaction barrier based on the four OER thermodynamic steps. The second term accounts for the dissipation and is related to the kinetics of the reaction. The following research focuses solely on the thermodynamics. Therefore, in this research, the limiting process that governs the required over-potential for the OER to evolve the OER will depend on the largest thermodynamic energy barrier.

The Gibbs energies of the each of the reactions provide a quantitative measure of the direction and required energy to produce a product from a group of reactants. In this study, the enthalpies of formation are determined through DFT calculations of the reactants and their constituents. By taking the difference in total energy between the reactants and their constituents an estimate for the enthalpies can be determined.  Because the oxygen evolution reaction involves four reaction steps, four Gibbs free energies were calculated. The four reaction steps, Equations~\plainref{eq:1}-\plainref{eq:4} can rearranged as follows,

\begin{equation}
\Delta G_1=\Delta G_{HO*}-\Delta G_{H_2O}-eU+k_BT ln(a_{H+})+ZPE,
\label{eq:5}
\end{equation}

\begin{equation}
\Delta G_2=\Delta G_{O*}-\Delta G_{HO*}-eU+k_BT ln(a_{H+})+ZPE,
\label{eq:6}
\end{equation}

\begin{equation}
\Delta G_3=\Delta G_{HOO*}-\Delta G_{O*}-eU+k_BT ln(a_{H+})+ZPE,
\label{eq:7}
\end{equation}

\begin{equation}
\Delta G_4=\Delta G_{O2}-\Delta G_{HOO*}-eU+k_BT ln(a_{H+})+ZPE,
\label{eq:8}
\end{equation}
here the $\Delta G$ is the Gibbs energies, $eU$ is the applied over-potential, $k_BT ln(a_{H+})$ is related to the temperature dependent entropy, and $ZPE$ is the zero-point energy. The entropy and zero-point energy for each of the reactants is tabulated in the supplemental material. The difference between the reactants and the products provides an estimate of the Gibbs free energy for each of the reaction steps. The asterisks symbol in Equation~\plainref{eq:5}-\plainref{eq:8} denotes the adsorption of the reactant on the MOF framework. 

\section{Results and Discussion}
The Gibbs free energy for each of the reactants and products were determined by relaxing the structure in each configuration and subtracting the total energy from the summation of the constitute atoms. A detailed discussion of how these Gibbs energies were calculated within the DFT framework is provided in the supplemental material. Table~\plainref{t:h} is the predicted Gibbs energies of the reaction for each of the radicals adsorbed on the MOF. The $\Delta G$ are associated with the four reactions of Equations~\plainref{eq:4}-\plainref{eq:8} were determined from the total energies from the DFT results. The columns of the table are associated with the three different linker designs. The smallest value in each of the rows is attributed with the most stable state. Of all of the designs, the hydroperoxyl ($HOO*$) case proves to be the most stable of all the products. Similarly, the hydroperoxyl binding is greatest on oxide surfaces. While this is results is not the whole picture of the OER, as that conclusion depends on the difference between reaction states, the difference between G$_{*}$ and G$_{HOO*}$ illustrates that there is a required potential to dissociate hydroperoxyl. Or, more concisely stated, the reaction will not be spontaneous.

\begin{table}
\begin{center}
  \begin{tabular}{ c | c | c | c }
    \hline \hline
    \multicolumn{3}{r}{~~~Gibbs Energy of Reaction (eV)} \\
    \cline{2-4}
    Product & BDC & BDC+NH$_2$ & BDC+NO$_2$ \\ \hline \hline
    $G_{*}$          & -836.3882918 & -909.4295651   & -3412.6696162\\ \hline
    $ G_{HO*}$    & -844.7804776 & -918.3767998   & -3421.6962788\\ \hline
    $G_{HOO*}$  & -849.5203220 & -922.7826306   & -3425.7845249\\ \hline
    $G_{O*}$       & -840.8630679 & -914.1245329   & -3417.2046487\\ \hline
    \hline
  \end{tabular}
\end{center}
\caption{DFT prediction of the Gibbs free energies of reaction for three MOF designs. The asterisk denotes the attachment of the radical to the MOF material. Note, these energy values are pseudo-potential specific.}
\label{t:h}
\end{table}
\begin{figure}
\centering
\includegraphics[width=0.6\columnwidth]{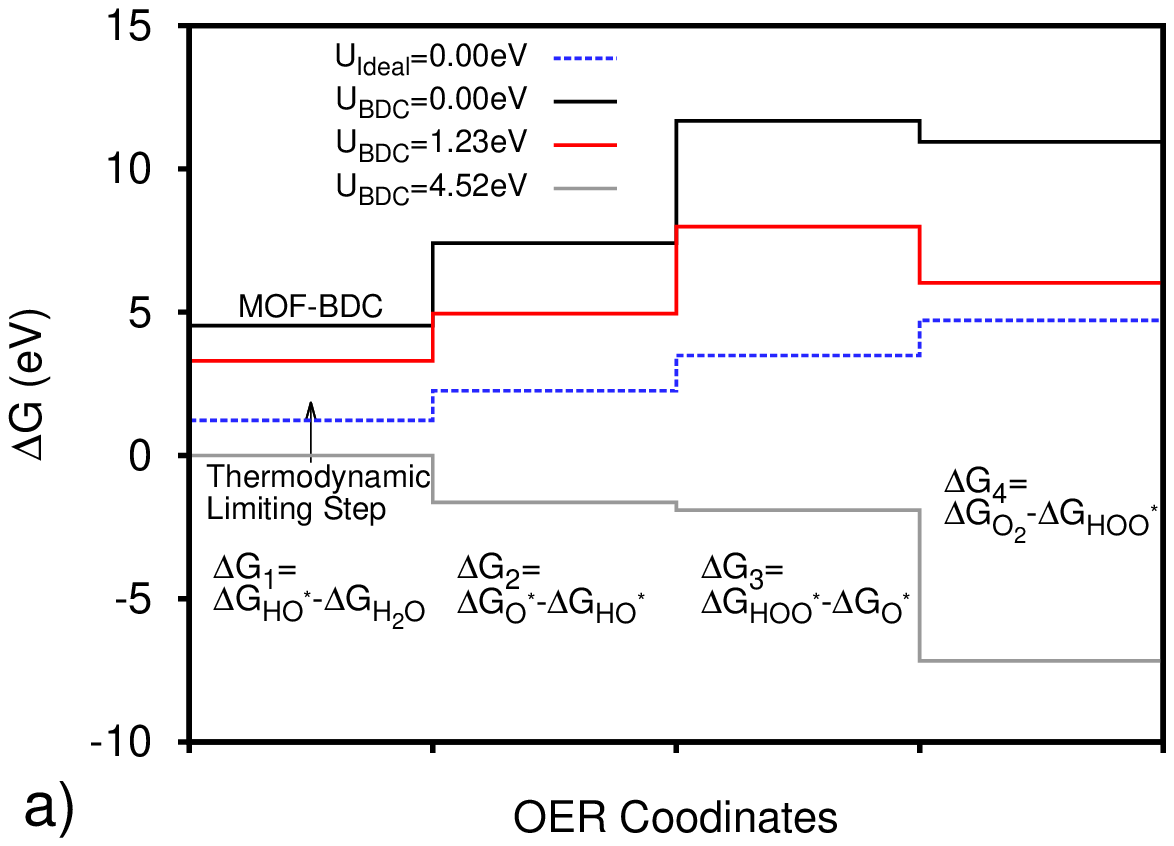}\\
\includegraphics[width=0.6\columnwidth]{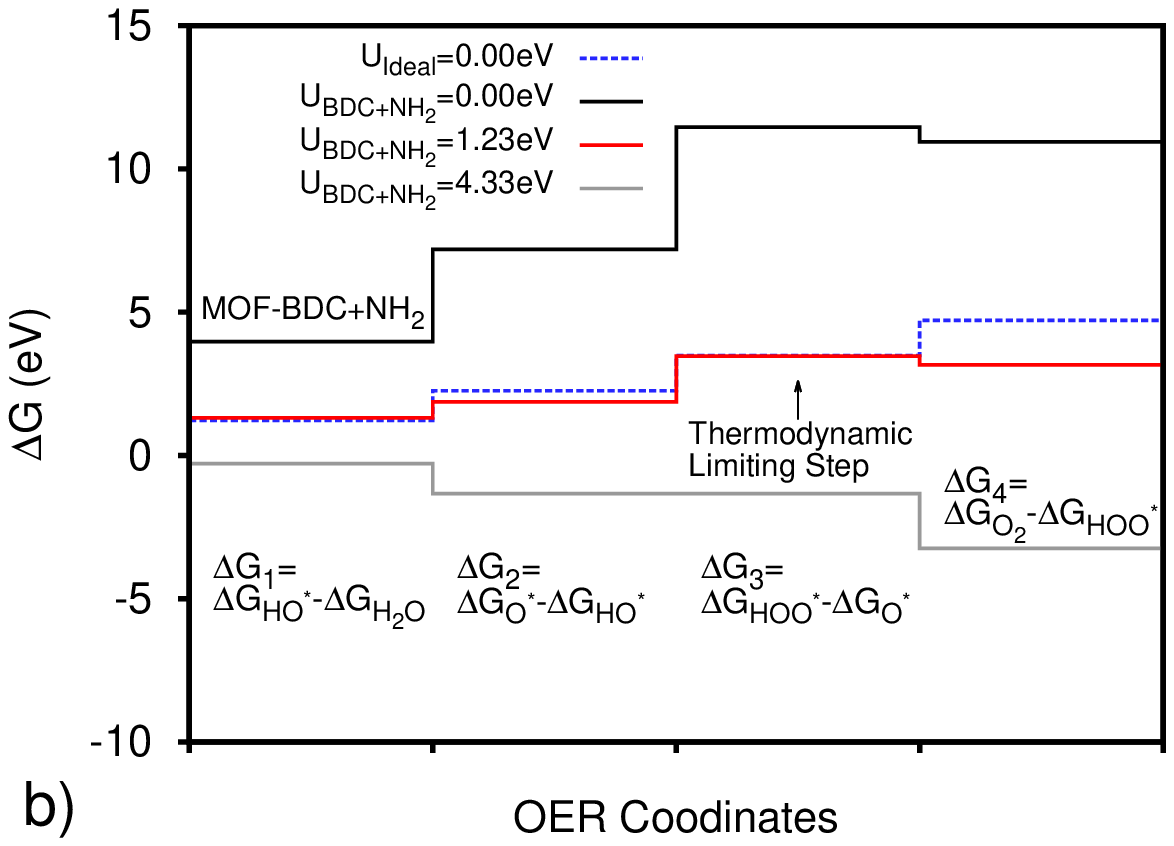}\\ 
\includegraphics[width=0.6\columnwidth]{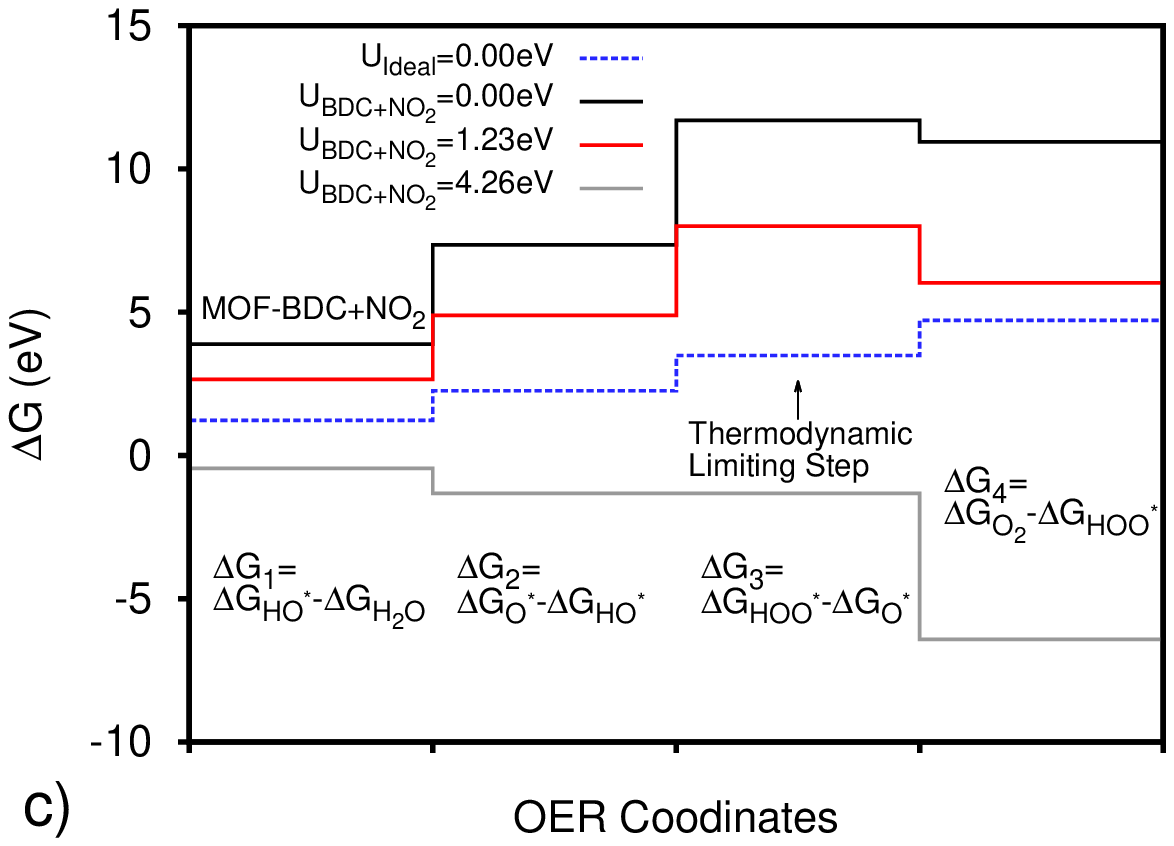} 
\caption{Gibbs energy for each of the linker designs for each of the four OER steps. The top figure (a) is for BDC, (b) is for BDC+NH$_2$, and (c) is for BDC+NH$_2$. The following plots correspond to the data in Table~\plainref{t:g}.}
\label{f:gibbs}
\end{figure}
Using the Gibbs energy for each of the products and reactants, the overall Gibbs energy of formation for each reactant steps (Equation~\plainref{eq:5}-\plainref{eq:8}) can be predicted. The results for each of the four reaction step are provide in Table~\plainref{t:g}. The Gibbs energies outlined in Table~\plainref{t:g} are visualized in Figure~\plainref{f:gibbs}. Note,  the reaction energies are compounded for each step. The ideal case is to have a reaction energy for each step that is proportional to 1.23V (see blue dashed line in Figure~\plainref{f:gibbs}, which is provided from the hydrogen electrode. The solid red line in the same figure is results of a 1.23V over-potential applied to the system, this in turn lowers the reaction barriers. However, for all the cases the Gibbs energy is still positive, indicating that there is still a reaction barrier. The gray line in Figure~\plainref{f:gibbs} is the potential required to overcome the highest reaction potential. The ideal case is to have the high reaction barrier be 1.23V. Therefore, when resulting Gibbs energy when a 1.23V potential is applied the Gibbs energies are zero and the reaction can spontaneously evolve.  

Focusing on the BDC MOF in Figure~\plainref{f:gibbs}, the thermodynamic limiting step is the first step where the water is dissociated into an adsorbed hydroxyl. For the BDC+NO$_2$, and BDC+NH$_2$ the thermodynamic limiting step was the third step that involves the product of absorbed hydroperoxyl from absorbed oxygen. The reason for the high barrier in the case of BDC+NO$_2$ and BDC+NH$_2$ MOF stems from nature of binding for both the oxygen and hydroperoxyl. By investigating the enthalpies for hydroperoxyl for all three MOFs as provided in Table~\plainref{t:h}, the hydroperoxyl proves to be the most stable (lowest relative energy). This means that the hydroperoxyl does not readily de-absorb without additional over-potential applied to the system or at least provides more resistance than the other reactions. Figure~\plainref{f:atomic_all}(a,d,g) is an atomic visualization of the hydroperoxyl binding with the aromatic carbon ring. Note, the orientation of the hydroperoxyl does vary, this is an initial indication that the functionalization of the MOF does influencing the binding and the orientation of the radical. Furthermore, the distance between the hydroperoxyl and the aromatic ring provides justification that the absorption is a physioabsorption type interaction involving secondary type bonding. The distance between the radical varied between 2.9 to 3.4\AA~ as illustrated in Figure~\plainref{f:atomic_all}.

\begin{table}
\begin{center}
  \begin{tabular}{ c | c | c | c }
    \hline \hline
    \multicolumn{3}{r}{~~~Gibbs Energy of Formation (eV)} \\
    \cline{2-4}
    Reaction & BDC & BDC+NH$_2$ & BDC+NO$_2$ \\ \hline \hline
    $\Delta G_{1}$ & 4.5269018 & 3.9718529 & 3.8924250\\ \hline
    $\Delta G_{2}$ & 2.8877472 & 3.2226043 & 3.4619676\\ \hline
    $\Delta G_{3}$ & 4.2618335 & 4.2609899 & 4.3392113\\ \hline
    $\Delta G_{4}$ & -0.7371574 & -0.5161220 & -0.7542788\\ \hline
    \hline
  \end{tabular}
\end{center}
\caption{Predictions of the four oxygen evolution reaction steps using DFT to calculate free energy of formation. See the supplemental material for example calculation of these formation energies.}
\label{t:g}
\end{table}
The second large reaction barrier for the third reaction as provided in Table~\plainref{t:g} for both the nitro- and amino- funcationalized MOF, is the binding of the oxygen. Table~\plainref{t:h} of the Gibbs energy for oxygen absorption (G$_{O*}$)  indicate that the second high energetics, just below the bare surface enthalpy (G$_{*}$). From the visualization of the oxygen adsorption as seen in Figure~\plainref{f:atomic_all}(c,f,i), the oxygen binding is more chemisoption. This is reasoned based on the distance between the aromatic ring and the oxygen. For both BDC and BDC+NH$_2$ the distance between the oxygen and the carbon is less than 2\AA. To provide a reference, diatomic oxygen has a distance of approximatly 1.2\AA.  Focusing in on Figure~\plainref{f:atomic_all}, it is interesting to note that the oxygen has preferential binding towards the end of the aromatic ring. In the case of the BDC (sub-figure c) and the amino (sub-figure d), the aromatic ring is slightly distorted indicating some additional interaction between the oxygen and neighboring molecules. This distortion is not apparent in the nitro-functional case and is reasoned to stem from charge distribution in the aromatic ring as a result of the functionalization. The electron acceptor nature of the nitro-group was confirmed in previous studies~\cite{musho14} where it was found that the electron exchange of the aromatic ring heavily influenced the optical properties. In the case of the nitro- functionalization, the aromatic ring becomes electron favorable and O$^-$ radical has less motivation to chemically adsorb on the aromatic ring and instead relies on secondary interaction. This can be visually confirmed by the distance between the oxygen and aromatic ring and the decreased directionality of the binding as seen in Figure~\plainref{f:atomic_all}(i). Because of the decreased binding of the oxygen on the pore surface for BDC+NO$_2$, the hydroperoxyl reaction (step 3 of OER) requires an increased over-potential (4.33V) to evolve the adsorbed oxygen (O$^*$) into hydroperoxyl (HOO$^*$). This is a significant finding because, 1) it confirms that an interaction exists between the functional group and the radicals, and 2) this confirms that the nature of the radical bonding either through secondary physisorption or stronger chemisoprtion, can be used as a mechanism to control or tailor the reaction barriers. In the case of the nitro-group, the decreased binding leads to increased over-potential to associated hydroperoxyl.

Figure~\plainref{f:atomic_all} provides some significant insight in to the binding nature of the radicals. The most significant utility of this figure is the confirmation that the radicals have preferential binding to different locations within the MOF pore. This is significant because it provide a mean to avoid physical limitation with the OER reactions at a single reaction site. Furthermore, this figure illustrates that that the functionalization does influence the radical. This is confirmed by a change in orientation, for example, a rotation of the hydroperoxyl and the increased atomic distances. 
\begin{figure}[!ht]
\centering
\includegraphics[width=1.0\columnwidth]{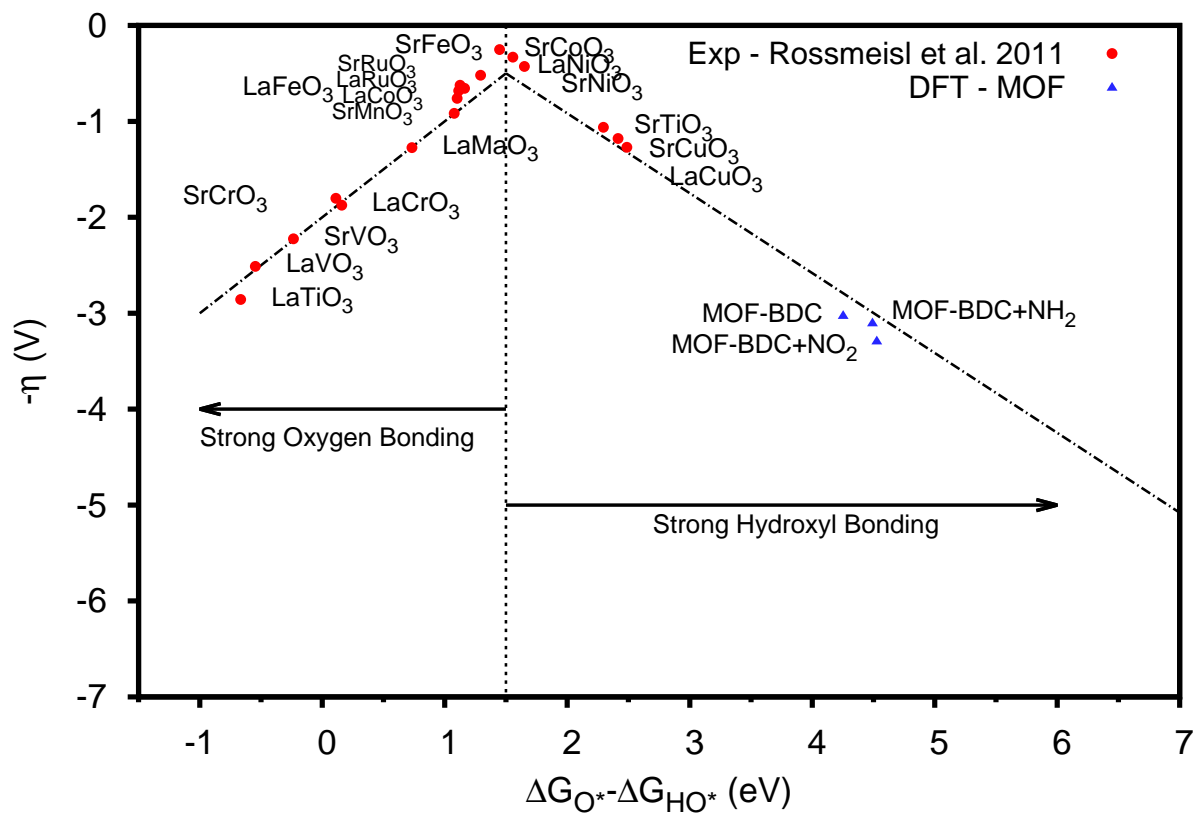} 
\caption{Volcano plot of the experimental values~\cite{rossmeisl11} and the DFT MOF results from this study. The optimal values are located near the apex. The MOF results indicate strong hydroxyl binding that fall near the volcano line because of the similar binding sites on the MOF. }
\label{f:volcano}
\end{figure}
\subsection{Volcano Plot of MOF Catalyst}
A useful metric for assessing the overall performance of a water oxidation activity of the potential catalyst is visualized through a volcano plot~\cite{trasatti80} that relates the difference in Gibbs energy to the required over-potential. The volcano plot for the three MOF designs along with a host of planar catalyst (source~\cite{rossmeisl11}) is provided in Figure~\plainref{f:volcano}. The utility of the volcano plot has been demonstrated for planar catalyst very successfully~\cite{trasatti80,rossmeisl11}. The y-axis is associated with the maximum over-potential, $\eta=max(\Delta G_{1},\Delta G_{2},\Delta G_{3},\Delta G_{4})-1.23$. The $1.23V$ is associated with the potential provided by the standard hydrogen electrode. The x-axis is the difference between the Gibbs energies, $\Delta G_{O^*}-\Delta G_{HO^*}$. The objective of the plot is to arrange a qualitative measure relative to other catalyst materials, given their oxygen and hydroxyl binding energy. The optimal position on the volcano plot is near the apex where the over-potential to overcome both the hydroxyl and oxygen surface absorption is minimized. In the case of the three MOF designs, all are located on the right portion of the volcano associated with strong hydroxyl bonding. At first glance, it is noted that these MOFs are not optimal for water oxidation, however, there are two significant finding that can be taken away from this plot. The first finding is that MOF oxidation follows the relationship of the volcano plot. This provides some confidance in these predictions. This also indicates that the oxygen and hydroxyl bonding are limiting reaction steps in the OER. Because the difference in these two Gibbs energies fall near the volcano lines, it is reasoned that the binding site at the same location on the MOF pore. This can be visually confirmed through Figure~\plainref{f:atomic_all}.The slope of the lines on the volcano correspond to a fundamental difference between the binding of hydroxol and oxygen to a surface. If a point falls far from these lines, the reader should check if the binding location is similar; this is not trivial for these organic MOF materials. The second finding from the volcano plot is that there is a confirmed degree of tailoribility through functionalization of the MOF. As noted in the previous section, the tailoribility is most significantly associated with the oxygen affinity, which is a function of carbons charge state. If the functionalization is able to modify the aromatic carbon's charge state, the binding of oxygen can be modified. This concept becomes even more difficult when considering the charge-transfer influences, which were not accounted for in this research.
\begin{figure}[!ht]
\centering
\includegraphics[width=1.0\columnwidth]{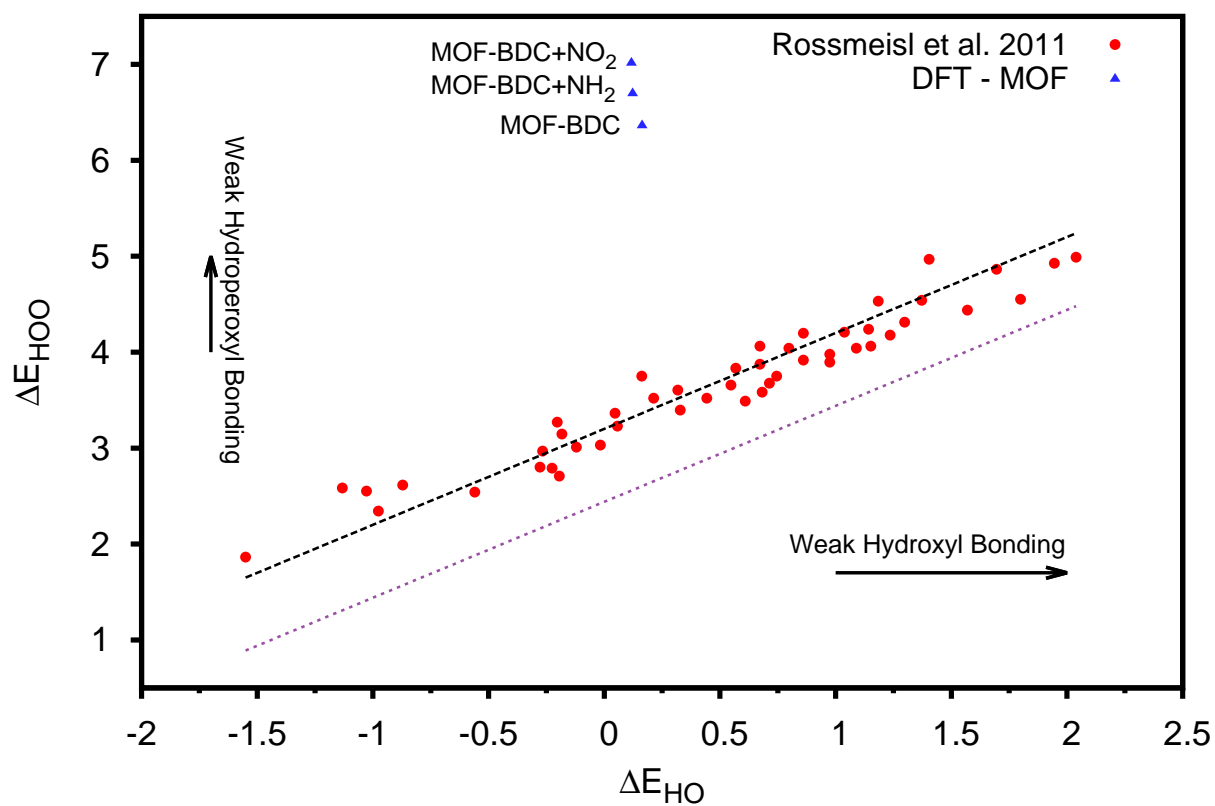} 
\caption{Plot of the absorption energy for experimental values~\cite{rossmeisl11} and DFT results from the three MOF designs. The black dashed line is associated with least square fit of the experimental data. The purple dash-dot is associated with the optimal line. The MOF do not fall on the linear lines because the binding spot is not the same for hydroperoxyl and hydroxyl.}
\label{f:adsorption}
\end{figure}
\subsection{Adsorption Energy.}
Another comparison that provides some confidence in the computational results and their association with planar catalyst thermodynamics related to the OER, is the prediction of the adsorption energy. It is well accepted that the adsorption energy of hydroxyl (HO) and hydroperoxyl (HOO) should fall on a line with a slope of unity. Rossmeisl~\cite{rossmeisl11} indicates that this has to do with the two preferring the same type of binding side and a fundamental constant associated with the binding energy. In the case of the MOF binding location, as illustrated in Figure~\plainref{f:atomic_all}(a,d,g) and Figure~\plainref{f:atomic_all}(b,e,h), the binding site is not at the same location and the location was found to depend on the type of radical (HO or HOO) and the functionalization. This is confirmed in Figure~\plainref{f:adsorption} of the adsorption energy with accompanying planar results. The reader should note that the black dashed line is a least square fit on the experimental data~\cite{rossmeisl11}. This fit still has a slope of unity but the y-intercept is 3.4eV. The dashed-dot line below the least square line outlines the optimal line of operation. For comparison the MOF results are shown in Figure~\plainref{f:adsorption} and prove to fall above the experimental results. Another trend that can be taken from this figure is the dependent of the hydroperoxyl adsorption with respect to the functionalization. This results in a vertical relative orientation of the three MOF materials within the plot. As mentioned in the previous section and visualized in Figure~\plainref{f:atomic_all}(a-c), the aromatic ring's charge state govern the binding of the radical. The MOF BDC+NO$_2$ proves to have the weakest of hydroperoxyl binding energy, however, the MOF BDC+NH$_2$ proves to require the least over-potential. The over-potential was a result of the weaker oxygen binding in the case of the amino-functionalization. The vertical alignment of the three designs was a result of the functionalization influencing the orientation and binding of the hydroperoxyl. Here the as-is BDC MOF proved to have the highest hydroperoxyl binding, however, this binding can be controlled by the functionalization, as demonstrated by the vertical orientation in Figure~\plainref{f:adsorption}.

\section{Conclusion}
The objective of this study was to the study the influence of the MOF functionalization on the thermodynamics of the oxygen evolution reaction. The funcationalization of the MOF was confirmed to influence the oxygen evolution reaction when compared to homogeneous catalyst surfaces. The most significant attribute of the MOF when compared to a planar homogeneous surface, is the heterogenetiy of binding location, which provides an avenue for overcoming limitations typically burdensome to homogeneous systems. The most significant of these is the ability to tailor the adsorption energy of hydroperoxyl binding, which proves burdensome with typical oxide surfaces (often limiting step in OER). It was determined that the oxygen binding was a critical step in the OER as a results of the moderate binding energy and visually illustrated strong interaction (small bond length). However, the funcationalization and charge state of the aromatic ring had the most influence on hydroperoxyl, which was the radical with the high binding. This provided insight that the functionalization of the MOF materials can readily be used to change the charge state of the aromatic ring and correspondingly control the hydroperoxyl binding. Granted the MOF designs selected for this study did not prove to outperform state-of-the-art oxide catalysts, but the findings from this study provide further justification for MOF based photocatalyst. Moreover, these findings demonstrate that DFT can be used to search the large design space of MOF based materials in hopes of finding an optical MOF design.

\newpage

\begin{suppinfo}
\subsection{Calculation of Gibbs Energies}
The approach used to predict the GIbbs energies relies calculating the total energies of the constitute atoms (C, H, O, etc.) and then the total energy of the ensamble (O*, OOH*, etc). Recall, the asterisk ($*$) indicated the attachment of the radical the the MOF structure. To demonstrate the complete calculation for the BDC MOF structure the first step is to determine the number of constitute atoms in the MOF structure. For the case of BDC primitive cell there are 28 hydrogen, 32 oxygen, 48 carbons, 0 nitrogen, 6 zirconium. The total number of atoms is 114. Next, for each of the constitute atoms the total energy is calculated by placing the atoms in a large box. For these calculation the box was cubic with a unit cell length of at least 18 bohr or approximately 9.5 angstroms. The symmetry was also turned off in these simulations and only one k-point was used at the gamma location. The result from these simulation are provided in Table~\plainref{t:ion_energy}. There is some interesting chemistry that can be taken away from these values. For example, dioxygen has a lower absolute energy that the sum of two monooxygen molecules. This is why atmospheric oxygen is mostly found as dioxygen (O$_2$).

\begin{table}[!ht]
\begin{center}
  \begin{tabular}{ c | c }
    \hline
    Reactant & DFT Total Energy (eV)\\ \hline \hline
    H            &      -12.52441816 \\
    HO          &   -446.57204650\\
    H$_2$    &   -879.15804381 \\
    O            &   -427.11295053 \\
    O$_2$   &    -862.70289993\\
    C            &    -145.87031968 \\
    N            &    -261.86782932 \\
    Zr           & -1341.40223691\\
    \hline \hline
  \end{tabular}
\end{center}
\caption{Total energies calculated using DFT of the constitute atoms. Note, these values are pseudo-potential dependent. }
\label{t:ion_energy}
\end{table}

The enthalpies of formation can be calculated by forming the macromolecules and calculating the total energy and then adding up the corresponding energy of the previously calculated constitute atoms in a big box and taking the difference. This enthalpy value should be comparable to experimental values. To determine the Gibbs energies the enthalpies are added to the experimentally determined zero point energies (ZPE) and entropies (TS). The ZPE and TS values are provided in Table~\plainref{t:zpe}. The calculated Gibbs energies are provided in the right most column of Table~\plainref{t:sole_energy}. The reader should be aware that there is a common issue with the over-prediction of the binding energy in dihydrogen. This is a result of the over analyticity of the DFT functionals and an over-prediction of the hydrogen bonding in dihydrogen. The dihydrogen values calculated have been provided in Table~\plainref{t:sole_energy} but the experimental values for the standard hydrogen electrode (SHE) were used to get a Gibbs energy for the hydrogen ion (H$^+$). The standard hydrogen potential is 4.44V at 25C. The reaction for the SHE is 2H$^+$+2e$^-$-->H$_2$(g). Using the DFT predictions for the Gibbs energy of H$_2$ the Gibbs energy of $H^+$ can be calculated. This in turn forces these calculation to be referenced to the SHE. The reader should note that the resulting potential for H$^+$ is 1.03V, which is slightly less than the accepted value of 1.23V. This difference corrects for the over-prediction of the H$_2$ binding.

\begin{table}[!ht]
\begin{center}
  \begin{tabular}{ c | c | c }
    \hline
    Reactant & Enthalpy (eV) & Gibbs Energy (eV)\\ \hline \hline
    H$_2$O        &   -13.94573852  & -13.94875007\\
    HO                 &    -6.93467780 & -7.01898065\\
    HO$_2$        &   -12.40772458 & -13.09080412\\
    O$_2$          &    -8.47699886 & -9.97885002\\
    H$_2$          &    -6.65538263 & -6.499325059\\
    1/2H$_2$ (SHE)     &             -              & -1.03\\
    \hline \hline
  \end{tabular}
\end{center}
\caption{Enthalapies for the macromolecules. An experimental value for the standard hydrogen electrode was used because of the over-estimate of the hydrogen binding in DFT calculations. }
\label{t:sole_energy}
\end{table}

The enthalpies are calculated for the MOF structure in a similar manner. The total energy of the MOF and the macromolecule are calculated using DFT. As stated in the narrative the approach was to place the macromolecule in the vincinity of the MOF pore and relax the all of the ions. Since this study was only interested in the thermodynamics and not the kinetics of the reaction pathway an transient approach such as a elastic nudge method, which relies on sequential image potentials was not necessary. 

Table~\plainref{t:mof_gibbs} is a summary of all of the energies. The first column is the lone ion energies, which is calculation by the summation of all the corresponding energies in Table~\plainref{t:sole_energy} for each ion in the structure. The second column is the total energy calculate from the DFT calculation of the relaxed structure. The third column is the difference between the first two columns and corresponds to the Gibbs energies. Here lies an approximated because the experimental ZPE and TS values were not known for the MOF structure. Therefore, ZPE and TS are assumed zero and the Gibbs energy is proportional the enthalpy of formation.

\begin{table*}[!ht]
\begin{center}
  \begin{tabular}{ c | c | c | c }
    \hline
    Reactant & DFT Ion Total Energy (eV) & DFT Total Energy (eV) & Gibbs Energy (eV)\\ \hline \hline
    MOF                &  -29068.48689174  & -29904.87518364 & -836.38829189\\
    MOF+HO*        &  -29508.1242604  & -30352.90473806 & -844.78047762\\
    MOF+HOO*     &  -29935.23721097  & -30784.75753299 & -849.52032201\\
    MOF+O*          &   -29495.59984228  & -30336.46291021 & -840.86306793\\
    \hline \hline
  \end{tabular}
\end{center}
\caption{Total energies and Gibbs energies for the BDC MOF structure. Note, the ZPE and TS were not known so the Gibbs energies are proportional to the enthalpy of formation energies.}
\label{t:mof_gibbs}
\end{table*}

Now that all of the Gibbs energies are know for all of the MOF structures and macromolecules the difference in Gibbs energies can be calculated for all four of the reactions. The list of reaction can be found in the narriative section. They consist of two half reactions. Recall, the difference in Gibbs energies ($\Delta$G) also known as the free energy of a reaction provides a quantitative measure of the magnitude and direction of the reaction. For reaction with a zero  $\Delta$G the reaction is at equilibrium, a $\Delta$G less than zero corresponds to a spontaneous reaction, and a $\Delta$G great than zero corresponds to a non-spontaneous reaction. 

The corresponding free energies of reaction are provided in Table~\plainref{t:free_energy} for all four of the reactions. Stepping through the first reaction, $\Delta$G$_1$=$\Delta$G$_{HO*}$-$\Delta$G$_{H_2O}$,  a water molecule is dissociated within the MOF pore resulting in the adhesion of a hydroxyl molecule to the MOF. The difference between these two represent the energy to undergo this first reaction. The term $\Delta$G$_{H_2O}$ is the summation of the Gibbs energy of the MOF (from Table~\plainref{t:mof_gibbs}) and the lone H$_2$O (from Table~\plainref{t:sole_energy}). Similarly, the $\Delta$G$_{HO*}$ is take from Table~\plainref{t:mof_gibbs}. The difference as seen in Table~\plainref{t:free_energy} is 3.10eV.

\begin{table}[!ht]
\begin{center}
  \begin{tabular}{ c | c  }
    \hline
    Reaction & Free Energy of Reaction, $\Delta$G (eV)\\ \hline \hline
    $\Delta$G$_1$ & 4.5269018\\
    $\Delta$G$_2$ & 2.8877472\\
    $\Delta$G$_3$ & 4.2618335\\
    $\Delta$G$_4$ & -0.7371574\\
    \hline \hline
  \end{tabular}
\end{center}
\caption{Total energies and Gibbs energies for the MOF structure. Note, the ZPE and TS were not known so the Gibbs energies are proportional to the enthalpy of formation energies.}
\label{t:free_energy}
\end{table}

When analyzing the absolute comparison of the Gibbs energies for each of the reaction step provided in Table~\plainref{t:free_energy} the reader should be careful. As discussed in the paper the over potential can be determined by evaluating the following expression $\eta=max(\Delta G_{1},\Delta G_{2},\Delta G_{3},\Delta G_{4})-1.23$.  To assemble the volcano plot the term $\Delta GO^*-\Delta GHO^*$ can be determined by taking the difference of the corresponding values found in Table~\plainref{t:mof_gibbs}.

\subsection{Entropies and Zero Point Energies}
The zero point energy is only important for the di-molecules. The zero point energies were determined from experiments of the resonance wavelength~\cite{irikura07}. This resonance wavelength was converted to an energy using the following relation, E=$\hbar \omega$, where $\omega$ is the wavelength. The entropy values were also taken from experimental values found in the literature. The corresponding zero point energy (ZPE) and Entropy are provided in Table~\plainref{t:zpe}

\begin{table}[!ht]
\begin{center}
  \begin{tabular}{ c | c | c }
    \hline
    Reactant & Entropy (Ry) & ZPE (Ry) \\ \hline \hline
    H                & 0.025149  & 0.000000 \\ \hline
    H$_2$       & 0.028649  & 0.040124 \\
    O$_2$       & 0.044974  & 0.014405  \\
    OH             & 0.040274  & 0.034075  \\
    H$^+$     & 0.023884  & 0.000000  \\
    H$_2$O(l) & 0.041398  & 0.041176 \\
    HOO          & 0.050226  & - \\
    \hline \hline
  \end{tabular}
\end{center}
\caption{Entropy and zero-point energies (ZPE) for select diatomic molecules~\cite{irikura07}.}
\label{t:zpe}
\end{table}
\end{suppinfo}

\bibliography{journal}

\end{document}